\documentclass[aps,prd,a4paper,nosuperscriptaddress,nofootinbib,showpacs,showkeys,twocolumn,amsfonts,amssymb,amsmath]{revtex4-2}
\usepackage{amssymb,latexsym}

\usepackage{bbold}
\usepackage{color}
\usepackage{amscd}
\usepackage{times}
\usepackage{epsfig}
\usepackage{psfrag}
\usepackage{graphicx}
\usepackage{amsthm,amsmath}
\usepackage{mathrsfs}
\usepackage[colorlinks=true, allcolors=blue]{hyperref}
\usepackage{leftindex}
\usepackage{mathtools}
\usepackage[dvipsnames]{xcolor}
\usepackage{physics}
\newcommand{\ii}{\mathrm{i}}
\newcommand{\ee}{\mathrm{e}}
\newcommand{\epspol}{\varepsilon}
\newcommand{\mb}[1]{\mathbf{#1}}
\renewcommand{\order}[1]{\mathcal{O}(#1)}

\begin{document}

\title[]{Graviton-mediated entanglement due to light bending from a quantum rotor}

\author{Dripto Biswas}
\author{Sougato Bose $^{1}$}
\author{Anupam Mazumdar$^{2}$}
\author{Marko Toro\v{s}$^{3}$}
\affiliation{ 
$^{1}$ Department of Physics and Astronomy, University College London, London WC1E 6BT, United Kingdom\\
$^{2}$ Van Swinderen Institute, University of Groningen, 9747 AG Groningen, The Netherlands\\
$^{3}$ Faculty of Mathematics and Physics, University of Ljubljana, Jadranska 19, SI-1000 Ljubljana, Slovenia
}

\begin{abstract}
One of the key tests of the quantum nature of gravity is to test whether the virtual mediator of gravity between matter and photon gives rise to the quantum light-bending phenomenon.  The off-shell degrees of freedom, involving the spin-2 and spin-0 components of graviton, reproduce the classical deviation of light rays, as well as have been predicted to generate entanglement between matter and photon. This paper explores the generation of entanglement due to the quantum gravitational interaction in an optomechanical setup with a quantum rotor and photon. The virtual exchange of a graviton provides entanglement between the photon degrees of freedom and the spatial position of the quantum rotor, with the rotational state affecting its magnitude. We analyze the case of a high spinning rotor, in an approximately classical state of angular momentum, and quantify its effect on the gravitationally induced entanglement between the photon and the position of the quantum rotor. We show that the difference in the linear entanglement entropies, of prograde-and-retrograde motion of the photon with respect to the quantum rotor, provide tangible observable consequences.

\end{abstract}

\maketitle

\section{Introduction}

The pivotal moment for general relativity was the detection of the deflection of light during the solar eclipse in 1919 \cite{Dyson:1920cwa}. Einstein's prediction from general relativity matched the data, with the distinction that Newtonian and Nordström's gravity could not explain the correct deflection angle due to the Sun's gravitational potential. This momentous experimental result has paved the way into a quantum domain, where one would quantize gravity around the Minkowski background as a low-energy effective field theory, see \cite{Gupta,Scadron:2007qd,Donoghue:1994dn}. In fact, in a perturbative quantum theory of gravity, it is possible to explain the angular deflection due to an off-shell graviton exchange between a massive object and a photon, see~\cite{Scadron:2007qd}. 

All these profound observations have led one group to propose an experiment to test the quantum nature of gravity in a lab first between two quantum superposed masses interacting via a graviton exchange: ``spin entanglement witness for quantum gravity"~\cite{ICTS,Bose:2017nin}, and also~\cite{Marletto:2017kzi}, and second via matter-graviton-photon interaction, while treating all three entities at par with quantum mechanics~\cite{Biswas:2022qto}, see also~\cite{Carney:2021vvt}. The primary idea is rooted in entanglement: two quantum systems can entangle if there exists a quantum mediator~\cite{Marshman:2019sne,Bose:2022uxe}; in the quantum gravity induced entanglement (QGEM) protocol that entails an exchange of a graviton between the two quantum superposed nanoparticles~\cite{Bose:2017nin}, see theory papers~\cite{Belenchia:2018szb,Kafri_2015,
Carney_2019,Danielson:2021egj,christodoulou2023locally,christodoulou2019possibility,Vinckers:2023grv,elahi2023probing}, and for experimental implementation~\cite{Bose:2017nin,Pedernales20_GM,Marshman:2021wyk,vandeKamp:2020rqh,
Chevalier:2020uvv,Krisnanda_2020,Elahi:2024dbb,Zhou:2024pdl}, for an experimental white paper, see~\cite{Bose:2025qns}. 
\footnote{In fact, a single matter-wave interferometer depicts all the scopes of quantum gravity if gravity and matter are treated at par at a quantum level~\cite{Toros:2020krn}. However, it is extremely hard to witness the entanglement between graviton and matter~\cite{Rufo:2024ulr}. Hence, one requires an experimental setup where two of the interferometers are kept adjacent to each other, and the graviton-mediated entanglement is witnessed by building correlations between the two matter interferometers~\cite{Bose:2017nin} or matter and photon degrees of freedom in~\cite{Biswas:2022qto,Dutta:2026pty}. }.

In \cite{Biswas:2022qto}, we studied the entanglement arising from the matter-photon interaction via graviton exchange, and in \cite{Dutta:2026pty}, we provided an entanglement witness protocol for this optomechanical experiment. The aim of this paper is to introduce rotation in the matter sector and to show how the linearised Hamiltonian naturally provides interactions among position, angular momentum, and photon degrees of freedom. To exemplify the importance of rotation, we consider a large angular momentum of a massive object and show that it is possible to witness an entanglement between prograde and retrograde motions of the matter with respect to the photon beam. Any rotating source also generatepolarisation-changingng terms that are relevant to helicity asymmetry and Skrotskii/Faraday rotation \cite{Kopeikin:2001dz,Guadagnini:2002xx,Bodenner:2003}. The present computation keeps the \emph{polarization-diagonal} $\order J$ piece that fits directly into the modified optomechanical Hamiltonian. In pa,st a single-photon source has
enabled the experimental exploration of
multi-mode interference and entanglement within
the context of quantum field theory in curved space-
time, see~\cite{Fink_2017,PhysRevLett.123.110401,Toros:2020,Toros:2022}, a regime which can be thought of as an approximation where gravity is treated classically. However, in this paper, the aim will be to treat gravity at the quantum level along with the matter.

In this paper, we will not consider the effects of dephasing or decoherence. These effects have been partially taken into account
in \cite{Biswas:2022qto} as far as the matter sector is concerned. A white noise analysis for the photon has also been taken into account while preparing the witness to observe the entanglement in \cite{Dutta:2026pty}. We will leave the details of how to induce rotation for a separate topic. Note that nanoparticle-controlled rotation has been performed in these experiments, see \cite{Biswas:2022qto}.

\section{Scattering amplitude of a quantum rotor and a photon due to graviton exchange}

Let us begin our discussion of the static Newtonian potential between matter and photon. The starting point is the tree-level amplitude between matter and photon mediated via an off-shell graviton. We use the metric signature: $(-,+,+,+)$, graviton coupling $f^2=8\pi G$, 
where $G$ is Newton's constant. In this paper, we are working in {\it natural units} ($c=\hbar =1$).
We take $k^\mu$ and $k'^\mu$ are the incoming and outgoing photon four-momenta. The momentum transfer is given by: $q^\mu=k^\mu-k'^\mu$, small-grazing-angle approximation $q^2\simeq 0$ in the numerator, with $1/(q^2+\ii\epsilon)$ kept intact in the propagator, where $\epsilon$ is the usual Feynman regularization prescription.
In the harmonic-gauge, the graviton propagator is given by~\cite{Scadron:2007qd,Biswas:2022qto}
\begin{equation}
    D_{\mu\nu\alpha\beta}(q)=\frac{\ii P_{\mu\nu\alpha\beta}}{q^2+\ii\epsilon},
\end{equation}
where $q$ is the four-momentum of the off-shell propagating graviton, with the standard projection operator, which contains both spin-2 and spin-0 components of the graviton, see~\cite{Biswas:2013kla,Biswas:2011ar,Marshman:2019sne}~\footnote{Note that in this paper the position, momentum, angular momentum are all operators, i.e. $\hat x,~ \hat p,~ \hat J$, and so are the creation, annihilation operators, $\hat a, ~\hat a^{\dagger},~ \cdots$. However, for brevity, we will not put the hat-sign in the expressions; instead, we will remind the readers about the $c$-numbered entities and operator-valued entities whenever there is any confusion.}~\cite{Scadron:2007qd,Donoghue:1994dn}:
\begin{equation}
P_{\mu\nu\alpha\beta}=\frac12\left(\eta_{\mu\alpha}\eta_{\nu\beta}+\eta_{\nu\alpha}\eta_{\mu\beta}-\eta_{\mu\nu}\eta_{\alpha\beta}\right).
    \label{eq:gravprop}
\end{equation}
In the notation of \cite{Biswas:2022qto}, the static tree-level covariant amplitude is given by 
\begin{alignat}{1}
\label{eq:A16static_clean}
    S^{\rm cov}_{fi} &=(-\ii)^2 4 f^2 p'_{\delta}p_{\sigma}
    \frac{\ii P^{\delta\sigma}{}_{\mu\nu}}{q^2+\ii\epsilon}
    \Bigl[k'_{\alpha}k^{\mu}\eta_{\beta}{}^{\nu}+k_{\beta}k'^{\mu}\eta_{\alpha}{}^{\nu} \nonumber\\
   &-\eta_{\alpha\beta}k'^{\mu}k^{\nu}-\frac12\eta^{\mu\nu}k_{\beta}k'_{\alpha}\Bigr]
    \epspol'^{*\beta}\epspol^{\alpha}\,\delta^4(P_i),
\end{alignat}
which simplifies to
\begin{equation}
\label{eq:A18static_clean}
    S^{\rm cov}_{fi}=-2\ii f^2 p'_{\delta}p_{\sigma}
    \left(\frac{-2k'^{\sigma}k^{\delta}}{q^2+\ii\epsilon}\right)
    \epspol'^*\!\cdot\epspol.
\end{equation}
The Born prescription used in \cite{Biswas:2022qto} maps this amplitude to the static potential:
\begin{equation}
    V_0(r)=-\frac{2GM\omega}{r}\,\epspol'^*\!\cdot\epspol,
\end{equation}
where $M$ is the central mass, $\omega$ is the photon frequency, and $\epspol,\epspol'$ are the initial and final polarisation vectors, see \cite{Biswas:2022qto}~\footnote{This potential yields the correct deflection angle as produced by classical general relativity, see the computation in \cite{Scadron:2007qd}.}. Here, note that $V_0(r)$ is a quantum operator by virtue of the position and the polarisation, i.e. $r,~\epsilon$, see~\cite{Bose:2022uxe,Biswas:2022qto}.
We now derive its leading correction linear in the source angular momentum $\mb J$. For a localised stationary spinning source, the appropriate external source is the standard pole-dipole stress tensor in the source's rest frame,
\begin{eqnarray}
\label{eq:pole_dipole_x_clean}
    T^{00}(\mb x)&=&M\,\delta^{(3)}(\mb x),~~~
    T^{0i}(\mb x)=-\frac12\epsilon^{ijk}J_j\partial_k\delta^{(3)}(\mb x),\nonumber \\
    T^{ij}(\mb x)&=&\order {J^2,\partial^2}.
\end{eqnarray}
The first term is the mass monopole; the second term is the current dipole responsible for frame dragging. Here $\epsilon^{ijk}$ is the three-dimensional Levi-Civita symbol with $\epsilon^{123}=+1$.
We define the momentum-space source tensor by
\begin{equation}
    \Theta^{\mu\nu}_{(J)}(\mb q)=\int \dd^3x\,\ee^{-\ii \mb q\cdot\mb x}T^{\mu\nu}(\mb x).
\end{equation}

Using Eq.~\eqref{eq:pole_dipole_x_clean} gives
\begin{eqnarray}
\label{eq:ThetaJ_clean}
    \Theta^{00}_{(J)}(\mb q)=M,
    \qquad
    \Theta^{0i}_{(J)}(\mb q)=\Theta^{i0}_{(J)}(\mb q)=\frac{\ii}{2}\epsilon^{ijk}q_jJ_k,
    \nonumber \\ 
    \Theta^{ij}_{(J)}(\mb q)=\order{q^2,J^2}.
\end{eqnarray}
This is the spinning-source replacement for the static source factor appearing in \eqref{eq:A18static_clean}. The natural spinning-source extension of the tree-level covariant amplitude is given by \eqref{eq:A16static_clean} is given by:
\begin{alignat}{1}
\label{eq:A16J_clean}
    S^{\rm cov}_{fi}[J]=&(-\ii)^2 4 f^2 \Theta^{(J)}_{\delta\sigma}(\mb q)
    \frac{\ii P^{\delta\sigma}{}_{\mu\nu}}{q^2+\ii\epsilon}
    \Bigl[k'_{\alpha}k^{\mu}\eta_{\beta}{}^{\nu}+k_{\beta}k'^{\mu}\eta_{\alpha}{}^{\nu}\nonumber\\
    &-\eta_{\alpha\beta}k'^{\mu}k^{\nu}-\frac12\eta^{\mu\nu}k_{\beta}k'_{\alpha}\Bigr]
    \epspol'^{*\beta}\epspol^{\alpha}\,\delta^4(P_i).
\end{alignat}
Using Eq. \eqref{eq:gravprop} and Eq. \eqref{eq:ThetaJ_clean} in Eq. \eqref{eq:A16J_clean}, we immediately obtain, 
\begin{equation}
\label{eq:A18J_clean}
    S^{\rm cov}_{fi}[J]=-2\ii f^2\Theta^{(J)}_{\delta\sigma}(\mb q)
    \left(\frac{-2k'^{\sigma}k^{\delta}}{q^2+\ii\epsilon}\right)
    \epspol'^*\!\cdot\epspol\,\delta^4(P_i)+S_{\rm flip},
\end{equation}
where $S_{\rm flip}$ denotes the polarization-changing $\order J$ terms (see Eqs. \eqref{eq:Sflip_def_clean} and \eqref{eq:Sflip_0i_clean}). We show further details in the next section and also show how $S_{\rm flip}$ can be set to zero depending on the direction of $\mb J$.
Substituting Eq.~\eqref{eq:ThetaJ_clean} and using the eikonal regime $k'^\mu\simeq k^\mu=(\omega,\omega{\mb n})$, where ${\mb n}=\mb k/\omega$ is the photon propagation direction, yields the polarization-diagonal kernel
\begin{eqnarray}
\label{eq:A18Jdiag_clean}
    &S^{\rm cov}_{fi,\mathrm{diag}}[J]=
    \nonumber \\  
    &-2\ii f^2\frac{-2}{q^2+\ii\epsilon}\left[M\omega^2+\frac{\ii\omega}{2}({\mb n}\times\mb q)\cdot\mb J\right]\epspol'^*\!\cdot\epspol\,\delta^4(P_i).
\end{eqnarray}
The new term is linear in $\mb J$, linear in $\mb q$, and odd under ${\mb n}\to-{\mb n}$.

For completeness, we now make explicit the quantity $S_{\rm flip}$ that appeared in Eq.~\eqref{eq:A18J_clean}.  By definition, $S_{\rm flip}$ is the part of the spinning-source matrix element that is \emph{not} proportional to the polarisation-preserving structure $\epspol'^*\!\cdot\epspol$.  Starting from Eq.~\eqref{eq:A16J_clean} and carrying out the same projector contraction as in Appendix A of \cite{Biswas:2022qto}, we get
\begin{equation}
\label{eq:Sflip_def_clean}
    S^{\rm cov}_{fi}[J]
    =
    S^{\rm cov}_{fi,\mathrm{diag}}[J]
    +
    S_{\rm flip},
\end{equation}
with
\begin{equation}
\label{eq:Sflip_exact_clean}
    S_{\rm flip}
    =
    (-\ii)^2 4 f^2
    \frac{\ii}{q^2+\ii\epsilon}\,
    \Theta^{(J)}_{\delta\sigma}(\mb q)\,
    \Delta^{\delta\sigma}_{\rm flip}\,
    \delta^4(P_i),
\end{equation}
where
\begin{eqnarray}
\label{eq:Dflip_clean}
    \Delta^{\delta\sigma}_{\rm flip}
    &=
    k^\delta \epspol'^{*\sigma}(k'\!\cdot\!\epspol)
    +
    k^\sigma \epspol'^{*\delta}(k'\!\cdot\!\epspol)
    +
    k'^\delta \epspol^\sigma (k\!\cdot\!\epspol'^*)
    +\nonumber \\ 
  &  k'^\sigma \epspol^\delta (k\!\cdot\!\epspol'^*).
\end{eqnarray}
Eq.~\eqref{eq:Dflip_clean} is the part of the projector-contracted amplitude that mixes polarisation states.  It vanishes identically for a static source, because then $\Theta^{0i}=0$ and only the diagonal mass-monopole term survives.

For a stationary spinning source, the only components relevant at linear order in $\mb J$ are given by Eq.~\eqref{eq:ThetaJ_clean}.
Using $q^0=0$, $\epspol^0=\epspol'^0=0$ \footnote{$k^0 = k'^0=0$ is simply a convenient gauge choice for the physical external photons; on the other hand, $q^0 = (k-k')^0$ denotes the energy transfer in the scattering process, which is taken to be zero in order to extract the Born potential for a stationary (time-independent) source potential.}, and keeping only the linear $\order J$ terms, Eq.~\eqref{eq:Sflip_exact_clean} reduces to
\begin{eqnarray}
\label{eq:Sflip_0i_clean}
    S_{\rm flip}
    =
    (-\ii)^2 4 f^2
    \frac{\ii}{q^2+\ii\epsilon}\,
    2\Theta^{(J)}_{0i}
    \left[
        \omega\,\epspol'^{*i}(k'\!\cdot\!\epspol) \right.
       +\nonumber \\
      \left.  \omega'\,\epspol^{i}(k\!\cdot\!\epspol'^*)
    \right]\delta^4(P_i),
\end{eqnarray}
where $\omega=k^0$ and $\omega'=k'^0$.
We now specialise to the physically relevant case in which the photon propagates in a plane and the source spin is perpendicular to that plane.  Choosing the photon plane to be the $xy$-plane, we take
\begin{eqnarray}
\label{eq:geom_flip_clean}
    \mb J = J {\mb z},
    \qquad
    \mb p=(p,0,0),
    \nonumber \\
    \mb k=(p\cos\theta,p\sin\theta,0),
    \qquad
    \mb q=\mb k-\mb p.
\end{eqnarray}
Note that all the above entities are operator-valued ones. A convenient linear-polarisation basis is given by:
\begin{eqnarray}
\label{eq:polbasis_flip_clean}
    \mb\epspol_\alpha^{(1)}=(0,1,0),~~
    \mb\epspol_\alpha^{(2)}&=(0,0,1),
    \nonumber \\
    \mb\epspol_\beta^{(1)}=(-\sin\theta,\cos\theta,0),~~    
    \mb\epspol_\beta^{(2)}&=(0,0,1).
\end{eqnarray}
Here, polarisation $(1)$ lies in the photon plane, while polarisation $(2)$ is orthogonal to that plane. With $\mb J=J{\mb z}$ and $\mb q$ confined to the $xy$-plane, Eq.~\eqref{eq:ThetaJ_clean} gives
\begin{equation}
\label{eq:Theta03zero_clean}
    \Theta^{(J)}_{0i}
    =
    -\frac{\ii J}{2}(q_y,-q_x,0),
    \qquad\Rightarrow\qquad
    \Theta^{(J)}_{03}=0.
\end{equation}
This simple observation is enough to show that the polarisation-changing terms vanish.  Indeed, consider the matrix element for the transition from the in-plane polarisation $(1)$ to the out-of-plane polarisation $(2)$:
\begin{equation}
\label{eq:Sflip21_clean}
    S_{\rm flip}^{21}
    \propto
    \Theta^{(J)}_{0i}
    \left[
        \omega\,\epspol_\beta^{(2)i}(k'\!\cdot\!\epspol_\alpha^{(1)})
        +
        \omega'\,\epspol_\alpha^{(1)i}(k\!\cdot\!\epspol_\beta^{(2)})
    \right].
\end{equation}
Since $\epspol_\beta^{(2)}={\mb z}$, the first term is proportional to $\Theta^{(J)}_{03}$ and therefore vanishes by Eq.~\eqref{eq:Theta03zero_clean}.  The second term vanishes because $k^\mu$ lies in the $xy$-plane whereas $\epspol_\beta^{(2)}$ points along ${\mb z}$, so
\begin{equation}
    k\!\cdot\!\epspol_\beta^{(2)}=0~~ \Longrightarrow S_{\rm flip}^{21}=0\,.
    \label{eq:Sflip21zero_clean}
\end{equation}
Likewise, for the transition from the out-of-plane polarization $(2)$ to the in-plane polarization $(1)$,
\begin{equation}
\label{eq:Sflip12_clean}
    S_{\rm flip}^{12}
    \propto
    \Theta^{(J)}_{0i}
    \left[
        \omega\,\epspol_\beta^{(1)i}(k'\!\cdot\!\epspol_\alpha^{(2)})
        +
        \omega'\,\epspol_\alpha^{(2)i}(k\!\cdot\!\epspol_\beta^{(1)})
    \right].
\end{equation}
Now $\epspol_\alpha^{(2)}={\mb z}$, so the first term vanishes because $k'^\mu$ lies in the $xy$-plane and therefore
\begin{equation}
    k'\!\cdot\!\epspol_\alpha^{(2)}=0.
\end{equation}
The second term is proportional to $\Theta^{(J)}_{03}$ and again vanishes by Eq.~\eqref{eq:Theta03zero_clean}. Therefore,
\begin{equation}
\label{eq:Sflip12zero_clean}
    S_{\rm flip}^{12}=0.
\end{equation}
Eqs.~\eqref{eq:Sflip21zero_clean} and \eqref{eq:Sflip12zero_clean} show that, to linear order in $\mb J$,
\begin{equation}
\label{eq:Sflipvanish_clean}
    \boxed{
    \mb J \perp \text{ photon plane}
    \quad\Longrightarrow\quad
    S_{\rm flip}=0
    }
    \qquad\text{at }\order J.
\end{equation}
Thus, when the source angular momentum is normal to the plane of photon propagation, the entire $\order J$ correction is polarisation-diagonal and is already contained in Eq.~\eqref{eq:VJ_clean}.  In particular, for the half-ring geometry used below, where the optical trajectory lies in the equatorial plane and $\mb J=J{\mb z}$, there is no additional polarisation-mixing contribution to the entanglement entropy at linear order in $J$.

As a cross-check, our analysis agrees with the small-angle polarisation matrix derived in \cite{Guadagnini:2002xx}. The authors computed the scattering amplitude of a photon and a spinning mass. The small-angle assumption amounted to setting the momentum transfer between the initial and final photon 3-momenta $(|\mb k_i - \mb k_f| = 2p\sin \beta/2$, where $\beta$ is the scattering angle between the incoming and outgoing photon). They obtained the scattering (amplitude) matrix:
\begin{align}
    A \propto \begin{bmatrix} \cos \frac{\beta}{2} - i 2p \frac{J_z}{M} \sin \frac{\beta}{2} & - 2p \frac{\tilde{J}}{M}\sin^2\frac{\beta}{2} \\ 2p \frac{\tilde{J}}{M}\sin^2\frac{\beta}{2} & \cos \frac{\beta}{2} - i 2p \frac{J_z}{M} \sin \frac{\beta}{2} \end{bmatrix}
\end{align}
where $J_z$ is the component of the angular momentum normal to the photon plane, which contributes only to the diagonal part of the amplitude, while the off-diagonal entries are controlled by the in-plane component of the angular momentum: $\tilde J = J_x \cos \frac{\beta}{2} + J_y \sin \frac{\beta}{2}$.

We now use the same Born prescription as in \cite{Biswas:2022qto}, that is, we compute the effective potential
\begin{equation}
\label{eq:Born_clean}
    V(\mb d)=\frac{1}{4M\omega}\int \frac{\dd^3q}{(2\pi)^3}S^{\rm cov}_{fi}(\mb q)\ee^{\ii\mb q\cdot\mb d}.
\end{equation}
To align the overall normalization with Appendix A of \cite{Biswas:2022qto}, we require that the $J\to0$ limit reproduce the static potential~\footnote{The relevant Fourier transforms are:
\begin{equation}
    \int \frac{\dd^3q}{(2\pi)^3}\frac{\ee^{\ii\mb q\cdot\mb d}}{\mb q^2}=\frac{1}{4\pi d},
    ~~~
    \int \frac{\dd^3q}{(2\pi)^3}\frac{\ii q_i\,\ee^{\ii\mb q\cdot\mb d}}{\mb q^2}=-\frac{d_i}{4\pi d^3}.
\end{equation}
}. Applying Eq.~\eqref{eq:Born_clean} to Eq.~\eqref{eq:A18Jdiag_clean} gives the leading polarization-diagonal potential
\begin{equation}
\label{eq:VJ_clean}
    V^{\mathrm{diag}}_J(\mb d,{\mb n})=-\frac{2GM\omega}{d}\,\epspol'^*\!\cdot\epspol+2G\omega\frac{\mb J\cdot(\mb d\times{\mb n})}{d^3}\,\epspol'^*\!\cdot\epspol+\order {J^2}.
\end{equation}
This is the main correction to the static potential. For the equatorial propagation around a source with $\mb J=J{\mb z}$, the $\order J$ term flips sign between co-rotating and counter-rotating trajectories, consistent with the weak-field Kerr deflection splitting discussed in \cite{Kopeikin:2001dz, Guadagnini:2002xx}.

\section{Optomechanical coupling}
\begin{figure}
    \centering
    \includegraphics[width=0.85\linewidth]{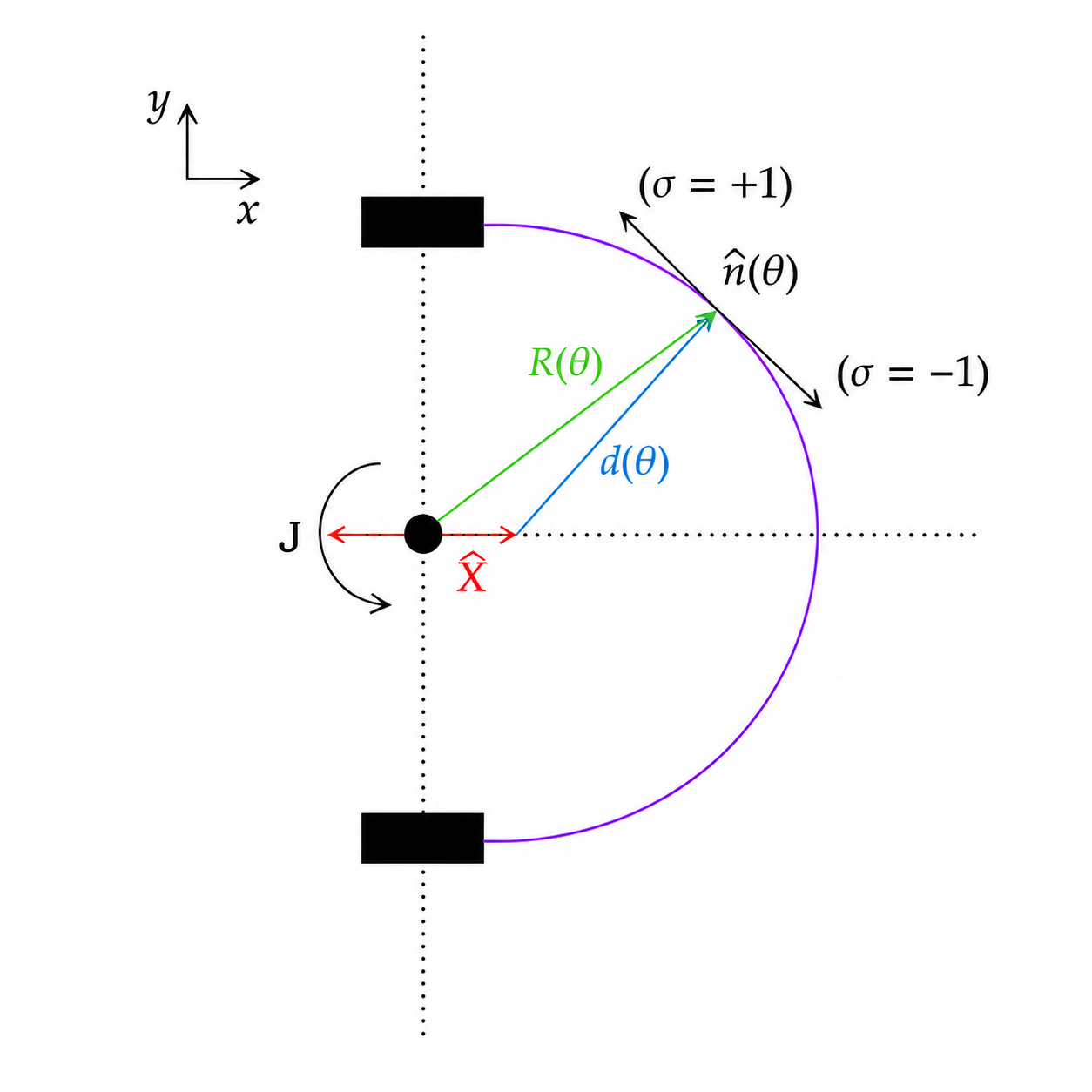}
    \caption{Schematic of the setup showing the various quantities defined in Sec. \ref{sec:geom_setup}. 
    At the centre there is a mass in a ground state initially $|0\rangle$, rotating 
    counterclockwise with an angular momentum ${\rm J}$.  The optical beam (purple) lies in the XY plane, and the tangential direction senses ($\sigma = \pm 1$), e.g, the photon beam can prograde and retrograde with respect to the rotation of the mass at the centre. We assume that the photon is in a coherent state $|\alpha\rangle$.}
    \label{fig:schematic_setup}
\end{figure}

\label{sec:geom_setup}

Let us now repeat the half-ring reduction \cite{Biswas:2022qto}. The optical path lies in the equatorial plane,
\begin{align}
    \mb R(\theta)&=(r\cos\theta,\,r\sin\theta,\,0), \nonumber \\
    {\mb X}&=(\delta x,\,0,\,0), \nonumber \\
    \mb d(\theta)&=\mb R(\theta)-{\mb X}, \\
    {\mb n}_{\sigma}(\theta)&=\sigma(-\sin\theta,\,\cos\theta,\,0),\qquad \sigma=\pm1, \nonumber 
\end{align}
with $\theta\in[-\pi/2,\pi/2]$ parameterizing the half-ring of radius $r$ and $\sigma$ labeling the two circulation senses. The static expansion used in \cite{Biswas:2022qto} is
\begin{equation}
    \frac{1}{|\mb d(\theta)|}=\frac{1}{r}+\frac{\delta x\cos\theta}{r^2}+\order{\delta x^2},
\end{equation}
\footnote{The half-ring integral turns $\cos\theta$ into the factor
\begin{equation}
    \int_{-\pi/2}^{\pi/2}\cos\theta\,\dd\theta=2.
\end{equation}
}
We now reproduce the static optomechanical interaction of \cite{Biswas:2022qto}. Using $\mb J=J{\mb z}$, we have
\begin{eqnarray}
    \mb J\cdot\bigl(\mb d(\theta)\times{\mb n}_{\sigma}(\theta)\bigr)=\sigma J\bigl(r-\delta x\cos\theta\bigr), \nonumber \\
    |\mb d(\theta)|=r-\delta x\cos\theta+\order{\delta x^2}.
\end{eqnarray}
Therefore,
\begin{equation}
\label{eq:spinexpand_clean}
    \frac{\mb J\cdot(\mb d\times{\mb n}_{\sigma})}{|\mb d|^3}
    =\sigma\frac{J}{r^2}\left(1+2\frac{\delta x\cos\theta}{r}\right)+\order{\delta x^2,J^2}.
\end{equation}
Integrating the linear term over the half-ring again produces the factor $2$, so the part linear in $\delta x$ is
\begin{equation}
    V^{(\sigma)}_{J,\mathrm{lin}}=+\frac{8\sigma GJ\omega}{r^3}\,\delta x.
\end{equation}
Combining this with the static linear term from \cite{Biswas:2022qto} gives the modified interaction
\begin{equation}
\label{eq:Vint_clean}
     V_{\mathrm{int}}^{(J,\sigma)}=-\left(\frac{4GM\omega}{r^2}-\frac{8\sigma GJ\omega}{r^3}\right)\delta x\, a^{\dagger} a.
\end{equation}
We now study the effective optomechanical coupling in presence of the rotation.Using the zero-point fluctuation operator,
\begin{equation}\label{eq:x_zpf_b_def_quantJ}
    \delta x=x_{\mathrm{zpf}}( b+ b^{\dagger}),
    \qquad
    x_{\mathrm{zpf}}=\frac{1}{\sqrt{2M\omega_m}},
\end{equation}
we obtain
\begin{equation}
\label{eq:g0J_clean}
     V_{\mathrm{int}}^{(J,\sigma)}=-g_{0,\sigma}^{(J)}( b+ b^{\dagger}) a^{\dagger} a,
\end{equation}
where
\begin{eqnarray}
\label{eq:g0J_formula_clean}
    g_{0,\sigma}^{(J)}&=&\left(\frac{4GM\omega}{r^2}-\frac{8\sigma GJ\omega}{r^3}\right)x_{\mathrm{zpf}}=g_0\left(1-2\sigma\frac{J}{Mr}\right),\nonumber \\
    g_0&=&\frac{4GM\omega}{r^2\sqrt{2M\omega_m}}\,,
\end{eqnarray}
$g_{0,\sigma}^{(J)}$ is the static single-photon coupling. Therefore, the explicit linear correction due to the rotation becomes:
\begin{equation}
\label{eq:deltag_clean}
    \delta g_{0,\sigma}=-2\sigma\frac{J}{Mr}\,g_0.
\end{equation}
Thus, the rotating source splits the two circulation directions linearly in $J$, where $\sigma =\pm 1$. The sign of the $\sigma$ label depends on which circulation direction is called ``$+$'' and on the orientation chosen for $\mb J$. The invariant statement is that the two directions split linearly in $J$. In the absence of $j$, we recover the results of \cite{Biswas:2022qto}.


\section{Linearised quantum Hamiltonian}

Note that 
${\mb r},~\delta  x,~{\mb J},~a_{\lambda'}^\dagger  a_\lambda$ are all operator valued entities.
In the polarisation-diagonal sector, and for the geometry in which the spin is perpendicular to the
photon plane, the polarisation matrix is proportional to the identity operator.  Therefore, the optical factor
reduces to the photon number operator
\[
     N_a
    =
    \sum_\lambda  a_\lambda^\dagger  a_\lambda .
\]
For a single occupied polarisation mode, this is simply $N_a=a^\dagger a$. The polarisation-diagonal potential is given by:
\begin{equation}
\label{eq:VJ_operator_quantJ}
    \boxed{
     V_J
    =
    \left[
        -\frac{2GM\omega}{ r}
        +
        2G\omega
        \frac{{\mb J}\cdot({\mb r}\times {\mb n})}{ r^3}
    \right] N_a .
    }
\end{equation}
Here, $M$ is still treated as the rest mass entering the monopole part of the potential, while
${\mb J}$ is now a quantum angular-momentum operator.

\subsection{Torsional oscillator representation of $ J_z$}

In the half-ring geometry used above, the optical path lies in the equatorial plane, and the spin is
taken to be normal to that plane:${\mb J}= J_z\,{\mb z}$.
Thus, only the component $ J_z$ enters the $\order{J}$ polarisation-diagonal correction.  To
write $ J_z$ in an oscillator basis, it is important to note that angular momentum is the
\emph{canonical momentum} conjugate to an angular coordinate. Therefore, the direct analogue of
$
    \delta x
    =
    x_{\rm zpf}( b+ b^\dagger)$,
is not $ J_z \propto ( c+ c^\dagger)$, but rather a momentum-quadrature expansion. Let $ \phi$ be a small angular displacement about the $z$ axis and let $\delta J_z$ be the
fluctuation of the angular momentum about a mean spin $J_0$:
\begin{equation}
\label{eq:J_split}
     J_z = J_0+\delta J_z .
\end{equation}
The canonical commutation relation is (in natural units):
\begin{equation}
\label{eq:phiJ_commutator}
    [\phi,\delta J_z]=i .
\end{equation}
For a torsional oscillator of moment of inertia $I$ and torsional frequency $\Omega_J$, the rotational
Hamiltonian becomes:
\begin{equation}
\label{eq:torsional_H}
     H_J
    =
    \frac{(\delta J_z)^2}{2I}
    +
    \frac12 I\Omega_J^2\phi^2 .
\end{equation}
Introducing a bosonic operator, $ c$, satisfying
$[ c, c^\dagger]=1$,
we define
\begin{equation}
\label{eq:phi_zpf_def}
    \phi
    =
    \phi_{\rm zpf}( c+ c^\dagger),
    \qquad
    \phi_{\rm zpf}
    =
    \sqrt{\frac{1}{2I\Omega_J}},
\end{equation}
and
\begin{equation}
\label{eq:J_zpf_def}
    \delta J_z
    =
    iJ_{\rm zpf}( c^\dagger- c),
    \qquad
    J_{\rm zpf}
    =
    \sqrt{\frac{I\Omega_J}{2}} .
\end{equation}
These definitions obey:
$\phi_{\rm zpf}J_{\rm zpf}={1}/{2}$,
and therefore reproduce Eq.~\eqref{eq:phiJ_commutator}.  With these definitions,
\begin{equation}
\label{eq:HJ_oscillator}
     H_J
    =
    \Omega_J
    \left(
         c^\dagger c+\frac12
    \right).
\end{equation}
Thus, the angular momentum operator becomes:
\begin{equation}
\label{eq:Jz_operator_c}
    \boxed{
     J_z
    =
    J_0+iJ_{\rm zpf}( c^\dagger- c).
    }
\end{equation}
where the above expression becomes the angular-momentum analogue of the mechanical displacement expansion
$\delta x=x_{\rm zpf}( b+ b^\dagger)$, with the crucial difference that $ J_z$ is a
momentum-like quadrature.

\subsection{Effective Hamiltonian}

For a photon
circulating in direction $\sigma=\pm1$, the spin-dependent part of the potential contains
\begin{equation}
\label{eq:spin_half_ring_operator_expansion}
    \frac{
         J_z\,
        \mb d(\theta)\times {\mb n}_\sigma(\theta)\cdot {\mb z}
    }{
        |\mb d(\theta)|^3
    }
    =
    \sigma\frac{ J_z}{r^2}
    \left(
        1+2\frac{\delta x\cos\theta}{r}
    \right)
    +
    \order{\delta x^2}.
\end{equation}
The term independent of $\delta x$ gives a spin-dependent optical phase,
\begin{equation}
\label{eq:spin_light_constant_term}
     V_{J,0}^{(\sigma)}
    =
    \chi_\sigma  J_z N_a ,
\end{equation}
where, with the same half-ring normalisation used above,
\begin{equation}
\label{eq:chi_sigma_def}
    \chi_\sigma
    =
    \frac{2\pi\sigma G\omega}{r^2}.
\end{equation}
If $J$ is treated as a classical number, this term is only a photon phase and can be removed.
However, since $ J_z$ is a quantum operator and the spin state is not an eigenstate of $ J_z$.
Then, Eq.~\eqref{eq:spin_light_constant_term} can entangle the optical mode with the rotational
degree of freedom. The term linear in $\delta x$ gives the modified optomechanical interaction
\begin{equation}
\label{eq:Vint_quantJ_before_zpf}
    \boxed{
     V_{\rm int}^{(J,\sigma)}
    =
    -
    \left(
        \frac{4GM\omega}{r^2}
        -
        \frac{8\sigma G\omega}{r^3} J_z
    \right)
    \delta x\, N_a .
    }
\end{equation}
Using Eq.~(\ref{eq:x_zpf_b_def_quantJ}),
we may write
\begin{equation}
\label{eq:Vint_quantJ_goperator}
     V_{\rm int}^{(J,\sigma)}
    =
    -
     g_{0,\sigma}^{(J)}
    ( b+ b^\dagger) N_a ,
\end{equation}
where the single-photon coupling is now itself an operator, where
\begin{equation}
\label{eq:g0_operator_quantJ}
    \boxed{
     g_{0,\sigma}^{(J)}
    =
    g_M
    -
    g_J J_z ,
    }
\end{equation}
with
\begin{equation}
\label{eq:gM_gJ_defs}
    g_M
    =
    \frac{4GM\omega}{r^2}x_{\rm zpf},
    \qquad
    g_J
    =
    \frac{8\sigma G\omega}{r^3}x_{\rm zpf}.
\end{equation}
Substituting Eq.~\eqref{eq:Jz_operator_c} gives
\begin{equation}
\label{eq:g0_operator_c_expanded}
     g_{0,\sigma}^{(J)}
    =
    g_M-g_JJ_0
    -
    i g_JJ_{\rm zpf}( c^\dagger- c).
\end{equation}
Therefore, the interaction becomes
\begin{align}
\label{eq:Vint_quantJ_expanded}
     V_{\rm int}^{(J,\sigma)}
    &=
    -
    \left(g_M-g_JJ_0\right)
    ( b+ b^\dagger) N_a
    \nonumber\\
    &\quad
    +
    i g_JJ_{\rm zpf}
    ( c^\dagger- c)
    ( b+ b^\dagger)
     N_a .
\end{align}
It is useful to define
\begin{equation}
\label{eq:g0cl_lambdaJ_defs}
    g_{0,\sigma}^{\rm cl}
    =
    g_M-g_JJ_0
    =
    \left(
        \frac{4GM\omega}{r^2}
        -
        \frac{8\sigma GJ_0\omega}{r^3}
    \right)x_{\rm zpf},
\end{equation}
and
\begin{equation}
\label{eq:lambdaJ_def}
    \lambda_{J,\sigma}
    =
    g_JJ_{\rm zpf}
    =
    \frac{8\sigma G\omega}{r^3}x_{\rm zpf}J_{\rm zpf}.
\end{equation}
Then
\begin{equation}
\label{eq:Vint_quantJ_final}
    \boxed{
     V_{\rm int}^{(J,\sigma)}
    =
    -
    g_{0,\sigma}^{\rm cl}
    ( b+ b^\dagger) N_a
    +
    i\lambda_{J,\sigma}
    ( c^\dagger- c)
    ( b+ b^\dagger)
     N_a .
    }
\end{equation}
The first term is the semiclassical spinning-source optomechanical interaction.  The second term is
new: it couples the photon number to both the centre-of-mass displacement quadrature and the
angular-momentum quadrature of the spinning source.
Including the displacement-independent spin-light term, the effective Hamiltonian becomes
\begin{align}
\label{eq:full_H_quantJ}
     H
    =
    \omega_m b^\dagger b
    +
    \Omega_J c^\dagger c
    &+
    \chi_\sigma J_z N_a
    -
    g_{0,\sigma}^{\rm cl}
    ( b+ b^\dagger) N_a \nonumber \\
    &+
    i\lambda_{J,\sigma}
    ( c^\dagger- c)
    ( b+ b^\dagger) N_a .
\end{align}
The zero-point term $\Omega_J/2$ has been omitted from Eq.~\eqref{eq:full_H_quantJ} because
it contributes only a global phase.


\section{Spin-eigenstate and classical limits}

If the rotational degree of freedom is prepared in a state sharply peaked around $J_0$, or if one
takes the formal classical limit
$    J_{\rm zpf}\rightarrow 0,~
     J_z\rightarrow J_0 $, then Eq.~\eqref{eq:Vint_quantJ_final} reduces to
\begin{equation}
\label{eq:Vint_classical_limit_quantJ}
     V_{\rm int}^{(J,\sigma)}
    \rightarrow
    -
    g_{0,\sigma}^{\rm cl}
    ( b+ b^\dagger) N_a ,
\end{equation}
which is precisely the semiclassical coupling used in the previous section. If, however, the spin state has appreciable quantum fluctuations in $ J_z$, the coupling
$ g_{0,\sigma}^{(J)}$ is operator-valued.  The optical field can then become entangled not only
with the centre-of-mass oscillator, $ (b,~b^{\dagger})$, but also with the rotational oscillator $ (c,~c^{\dagger})$.

We now compute the photon--matter entanglement generated by the spin-corrected
optomechanical Hamiltonian. In this section we specialise to the case in which
the angular momentum is either classical, \(J=J_0\), or the rotational degree of
freedom is prepared in a sufficiently sharp eigenstate of \( J_z\). In this
limit, the coupling \(g^{(J)}_{0,\sigma}\) is a c-number and the calculation is a
direct generalisation of the non-rotating case.
The relevant spin-corrected optomechanical interaction becomes:
\begin{equation}
     V_{\rm int}^{(J,\sigma)}
    =
    -g^{(J)}_{0,\sigma}
    ( b+ b^\dagger) a^\dagger a,
\end{equation}
where 
\begin{equation}
    g^{(J)}_{0,\sigma}
    =
    \left(
        \frac{4GM\omega}{r^2}
        -
        \frac{8\sigma GJ\omega}{r^3}
    \right)x_{\rm zpf}
    =
    g_0
    \left(
        1-2\sigma\frac{J}{Mr}
    \right).
\end{equation}
Here \(\sigma=\pm1\) labels the two optical circulation directions.  The
static coupling is $g_0
    =({4GM\omega}/{r^2})x_{\rm zpf}$ and $
    x_{\rm zpf}
    ={1}/{\sqrt{2M\omega_m}}$ in the natural units.
The full Hamiltonian for the centre-of-mass oscillator and the optical mode then becomes:
therefore
\begin{equation}\label{effH}
     H_\sigma
    =
    \omega_m b^\dagger b
    -
    g^{(J)}_{0,\sigma}
    ( b+ b^\dagger) a^\dagger a.
\end{equation}
For classical \(J\), the displacement-independent part of the spin potential
only gives an optical phase proportional to \( a^\dagger a\), and hence
does not contribute to the photon--matter entanglement.  If \(J_z\) is kept as
an operator, this statement no longer holds because the corresponding phase can
entangle the optical mode with the rotational degree of freedom. The photon-number operator
$N_a= a^\dagger a$
commutes with \( H_\sigma\).  Hence, the optical number operator evolves
independently.  We denote the photon-number indices by \(n\) and \(\ell\) to
avoid confusion with the source mass \(M\), and define
\begin{equation}
    G^{(J)}_\sigma
    =
    \frac{g^{(J)}_{0,\sigma}}{\omega_m},
    \qquad
    t=\omega_m\tau,
    \qquad
    \Lambda=|\alpha|^2 ,
\end{equation}
where \(\omega_m\) is the mechanical frequency, \(\tau\) is the laboratory
interaction time, and \(\alpha\) is the coherent-state amplitude of the optical
mode.

\section{Modified entanglement entropy}

Taking the initial state of the photon to be in a coherent state and the matter to be in a ground state, i.e.
\begin{equation}
    |\Psi(0)\rangle
    =
    |0\rangle_b\otimes|\alpha\rangle_a ,
    \qquad
    |\alpha\rangle_a
    =
    e^{-\Lambda/2}
    \sum_{n=0}^{\infty}
    \frac{\alpha^n}{\sqrt{n!}}|n\rangle_a ,
\end{equation}
we may write the evolved state under the Hamiltonian Eq.~(\ref{effH}), given by as
\begin{equation}
    |\Psi(t)\rangle
    =
    e^{-\Lambda/2}
    \sum_{n=0}^{\infty}
    \frac{\alpha^n}{\sqrt{n!}}
    |\phi_n(t)\rangle_b
    |n\rangle_a .
\end{equation}
For a fixed photon number \(n\), the oscillator evolves under Eq.~(\ref{effH}), and by completing the square in the Hamiltonian Eq.~(\ref{effH}), gives
\begin{equation}
     H_{\sigma,n}
    =
    \omega_m
    \left(
         b^\dagger-G^{(J)}_\sigma n
    \right)
    \left(
         b-G^{(J)}_\sigma n
    \right)
    -
    \omega_m
    \left(G^{(J)}_\sigma n\right)^2 .
\end{equation}
Starting from the oscillator ground state, the mechanical state in the
\(n\)-photon branch is therefore a coherent state,
\begin{equation}
    |\phi_n(t)\rangle_b
    =
    \exp\!\left[
        i\left(G^{(J)}_\sigma\right)^2n^2(t-\sin t)
    \right]
    \left|
        G^{(J)}_\sigma n(1-e^{-it})
    \right\rangle_b .
\end{equation}
Thus, different photon-number sectors displace the mechanical oscillator by
different amounts.  This is the origin of the photon--matter entanglement. The reduced state of the mechanical oscillator is obtained by tracing over the
optical mode:
\begin{equation}
    \rho_b(t)
    =
    {\rm Tr}_a|\Psi(t)\rangle\langle\Psi(t)|
    =
    e^{-\Lambda}
    \sum_{n=0}^{\infty}
    \frac{\Lambda^n}{n!}
    |\phi_n(t)\rangle\langle\phi_n(t)|.
\end{equation}
The linear entropy is defined as,
\begin{equation}
    S_{J,\sigma}(t)
    =
    1-{\rm Tr}_b\rho_b^2(t).
\end{equation}
Using the above form of \(\rho_b\), the purity is
\begin{equation}
    {\rm Tr}_b\rho_b^2(t)
    =
    e^{-2\Lambda}
    \sum_{n=0}^{\infty}
    \sum_{\ell=0}^{\infty}
    \frac{\Lambda^{n+\ell}}{n!\,\ell!}
    \left|
        \langle\phi_\ell(t)|\phi_n(t)\rangle
    \right|^2 .
\end{equation}
For coherent states,
\begin{equation}
    |\langle\beta|\gamma\rangle|^2
    =
    \exp[-|\beta-\gamma|^2].
\end{equation}
Here, the coherent-state displacement in the \(n\)-photon branch is
\begin{equation}
    \beta_n(t)
    =
    G^{(J)}_\sigma n(1-e^{-it}).
\end{equation}
Therefore,
\begin{align}
    |\beta_n(t)-\beta_\ell(t)|^2
    &=
    \left(G^{(J)}_\sigma\right)^2(n-\ell)^2
    |1-e^{-it}|^2
    \nonumber\\
    &=
    2\left(G^{(J)}_\sigma\right)^2(n-\ell)^2(1-\cos t).
\end{align}
Thus, we get
\begin{equation}
    \left|
        \langle\phi_\ell(t)|\phi_n(t)\rangle
    \right|^2
    =
    \exp\!\left[
        2\left(G^{(J)}_\sigma\right)^2
        (\ell-n)^2(\cos t-1)
    \right].
\end{equation}
The exact analogue of Eq. (21) of \cite{Biswas:2022qto} is then
\begin{align}
    S_{J,\sigma}(t)
    =
    1
    -
    e^{-2|\alpha|^2}
    \sum_{n=0}^{\infty}
    \sum_{\ell=0}^{\infty}
    \frac{|\alpha|^{2(n+\ell)}}{n!\ell!} \nonumber \\
    \times \exp\!\left[
        2
        \left(
            G^{(J)}_\sigma
        \right)^2
        (\ell-n)^2
        (\cos t-1)
    \right].
\end{align}

\subsection{Gaussian approximation for large \(|\alpha|\)}

For a bright coherent optical field, \(|\alpha|^2=\Lambda\gg1\), the Poisson
distribution may be approximated by a Gaussian,
\begin{equation}
    e^{-\Lambda}
    \frac{\Lambda^n}{n!}
    \simeq
    \frac{1}{\sqrt{2\pi\Lambda}}
    \exp\!\left[
        -\frac{(n-\Lambda)^2}{2\Lambda}
    \right].
\end{equation}
The double sum in the exact expression for \(S_{J,\sigma}\) can then be
approximated by a two-dimensional Gaussian integral.  Since the overlap depends
only on the photon-number difference \(n-\ell\), one introduces
\[
    u=\frac{n+\ell}{2},
    \qquad
    v=n-\ell .
\]
The \(u\)-integral is elementary, while \(v\) is distributed with variance
\(2\Lambda\).  This gives
\begin{equation}
    {\rm Tr}\,\rho_b^2(t)
    \simeq
    \frac{1}{
    \sqrt{
        1
        +
        8
        \left(
            G^{(J)}_\sigma
        \right)^2
        |\alpha|^2
        (1-\cos t)
    }} .
\end{equation}
Using the same Gaussian approximation to the Poisson weights as in \cite{Biswas:2022qto}
yields
\begin{equation}\label{lin-ent}
    S_{J,\sigma}(t)
    \approx
    1
    -
    \frac{1}{
    \sqrt{
        1
        +
        8
        \left(
            G^{(J)}_\sigma
        \right)^2
        |\alpha|^2(1-\cos t)
    }} .
\end{equation}
In the weak-coupling regime
\[
    \left(
        G^{(J)}_\sigma
    \right)^2
    |\alpha|^2
    \ll1,
\]
this becomes
\begin{equation}
    S_{J,\sigma}(t)
    \approx
    S_{\max,J,\sigma}(1-\cos t),
    \qquad
    S_{\max,J,\sigma}
    =
    4
    \frac{
        \left(
            g^{(J)}_{0,\sigma}
        \right)^2
        |\alpha|^2
    }{
        \omega_m^2
    } .
\end{equation}
Comparing with the static (non-rotating) case gives
\begin{equation}
    S_{\max,J,\sigma}
    =
    S_{\max,0}
    \left(
        1-2\sigma\frac{J}{Mr}
    \right)^2
    \approx
    S_{\max,0}
    \left(
        1-4\sigma\frac{J}{Mr}
    \right),
\end{equation}
where
\begin{equation}
    S_{\max,0}
    =
    4
    \frac{g_0^2|\alpha|^2}{\omega_m^2}
\end{equation}
is the static result.

\subsection{Short-time limit}

Define the light-enhanced coupling
$g^{(J)}_\sigma = g^{(J)}_{0,\sigma}|\alpha|$, and in the short-time regime \(t = \omega_m\tau \ll 1\), we can reduce Eq.~(\ref{lin-ent}):
\begin{equation}
S^{\rm short}_{J,\sigma}
=
2\left(g^{(J)}_\sigma\right)^2\tau^2
=
2\left(g^{(J)}_{0,\sigma}\right)^2|\alpha|^2\tau^2 .
\end{equation}
Using Eq. (42) in the limit, $t = \omega_m\tau \ll 1$, yields:
\begin{equation}
S^{\rm short}_{J,\sigma}
=
S^{\rm short}_0
\left(
1-2\sigma\frac{J}{Mr}
\right)^2
\approx
S^{\rm short}_0
\left(
1-4\sigma\frac{J}{Mr}
\right),
\end{equation}
where
\begin{equation}
S^{\rm short}_0
=
2g_0^2|\alpha|^2\tau^2
=
\frac{4G^2M\omega^2|\alpha|^2\tau^2}{r^4\omega_m}.
\end{equation}
where the light intensity \(I\) can be related to the occupation number of photons
\(|\alpha|\), the electric field \(E_c\) can be related to the cavity volume
\(V_c=(\pi r)(\pi w_d^2)\), where \(w_d\) is the cavity waist, and \(r\) is the
radial distance of the photon cavity from the massive object.
\begin{equation}
|\alpha|^2
=
\frac{2I}{\epsilon_0E_c^2},
\qquad
E_c
=
\sqrt{\frac{\omega}{2\epsilon_0V_c}},
\qquad
|\alpha|
=
\sqrt{\frac{4IV_c}{\omega}} .
\end{equation}
In \cite{Biswas:2022qto}, we set the following values for the mechanical oscillator,
\(M=10{\rm kg}\), \(\omega_m=2\pi\times150{\rm Hz}\) (for these values and
\(\rho\sim10^4{\rm kgm}^{-3}\), corresponding to radius \(R\sim6{\rm cm}\),
\(I=10^{13}{\rm Wcm}^{-2}\) at the optical wavelength
\(\lambda=1\mu{\rm m}\) \((\omega=2\pi/\lambda)\), while for the half-ring cavity
we set the radius to \(r=25{\rm cm}\) and the waist to \(w_d=6{\rm cm}\), such
that the total power circulating in the cavity is \(\sim1{\rm PW}\), which
corresponds to \(|\alpha|\sim10^{13}\). One can obtain
\(S^{\rm short}_0\sim O(1)\) for \(\tau\sim1{\rm ms}\) while squeezing the matter
oscillator \(\Delta x\sim x_{\rm zpf}e^\xi\sim6{\rm cm}\), for \(\xi\sim41\),
see \cite{Biswas:2022qto}. Recently, a $10$kg mirror has been motionally cooled close to the quantum ground state~\cite{Whittle_2021,vovrosh2017parametric}. Of course, squeezing all the way to $41$ is an extremely challenging task, and it will take many technical hurdles to be overcome. Also, petawatt laser beams are available~\cite{Yoon:2021}. However, these are on-shot lasers and running them continuously for $10$ms will also be a huge technical challenge.

However, with the rotation of the matter oscillator, there is an extra vantage that
we can study the difference of short-time entanglement (maximum values)
generated in the two cases (pro and retrograde motion of the photons), which
isolates the source spin directly,
\begin{equation}
S^{\rm short}_{J,+}-S^{\rm short}_{J,-}
\approx
-8\frac{J}{Mr}S^{\rm short}_0 .
\end{equation}
The difference is proportional to \(J/(Mcr)\) if we insert the units properly from the natural units.
The angular momentum of any object depends on its breaking point, the stress it
can handle. The largest stress the material can handle without breaking is given
by \(\sigma\sim\rho v^2\), while \(J/(Mc)\sim\kappa Rv/c\), where
\(\kappa=2/5\) for a rigid sphere, and $R$ is the radius of the sphere. Hence, the maximum velocity for a sphere can be written as:
\(v_{\max}\sim\sqrt{\sigma_{\max}/\rho}\), for a dense object with
\(\rho\sim2.2\times10^4{\rm Kgm}^{-3}\), such as Iridium of mass, \(M=10{\rm kg}\),
\(R=4.72{\rm cm}\), \(\sigma_{\max}\sim 10^9{\rm Pa}\), which gives
\(v_{\max}/c\sim 10^{-5}\), hence \(J/(Mcr)\sim O(10^{-7})\) for $r=25$ cm (the c.o.m distance between the spinning sphere and the laser beam). The current lab results for,  $(J/Mcr)\sim 12\times 10^{-10}$~\cite{Schuck:2018}, and $\sim 10^{-6}$~\cite{Ahn:2018}, but for much smaller masses, while keeping $r=25$cm.
Although in our case the difference between retrograde and prograde linear entropies is small, it could be detectable in future experiments if decoherence parameters are well controlled. Typically, for a small entanglement phase, there exists a relationship: $\phi_{ent}\propto S_{J,\sigma}^{1/2}$~\cite{Marshman:2019sne}, which suggests that a small variation in the entanglement phase is a measurable quantity in a laboratory. In our case, the difference in the linear entropy will entail an entanglement phase difference of $\phi_{ent}\sim {\cal O}(10^{-3})$. 

Detecting such a small difference in the entanglement phase will depend on the witnessing scheme, and we may need to seek a new one for entanglement between a spinning sphere and a laser beam.
Another important facet will be the estimation of decoherence in a massive matterwave setup~\cite{Hornberger:2008xkz,
ORI11_GM,Schlosshauer:2019ewh,Biswas:2022qto,Schut:2024lgp,Tilly:2021qef},  and optical setup~\cite{Preskill-noise,caves1981quantum}. Besides these sources of decoherence, there are other sources of dephasing due to random acceleration noise and gravity gradient noise~\cite {Saulson:1984,Thorne_1999, Hughes_1998,Toros:2020dbf}, which need to be considered for such a precision-led experiment. However, these issues are beyond the current scope of the paper, as they depend on the choice of platform for creating a spatial superposition of matter and on the details of the witnessing mechanism.

In any case, our study provides the first-ever theoretical handle on rotational dynamics in the presence of a quantum gravitational interaction with a photon in a lab. Moreover, it will ascertain how a massless graviton can entangle a rotating quantum matter with a photon via the off-shell graviton exchange.

\section{Conclusion}

The highlight of the paper is to show that by including the spinning object, at a quantum level, the scattering cross-section between the former and a photon via a massless graviton exchange yields a polarisation-changing contribution, which is a known classical effect, see~\cite{Kopeikin:2001dz,Guadagnini:2002xx}.
However, demonstrating this effect of rotation at a quantum level in a laboratory will be very interesting and require a gyroscopically stable rotor, see~\cite{Ahn:2018,
Stickler:2021dho,Zhou:2024pdl,Perdriat:2024xiy,Perdriat:2024zje}. 
By imparting a classical angular momentum, the effect of spin can be witnessed in the entanglement entropy for the choice of prograde and retrograde motion of the photon with respect to the spinning object. The difference in the entanglement entropy is proportional to the angular momentum $\propto J/(Mcr)$. Further improvements in the observability of the entanglement can be made by squeezing the matter sector to enhance the superposition size. If we set $m = 100$ g (radius $R = 1$cm and density $\rho=10~{\rm g cm^{-3}}$, we can generate unit entanglement $S_0^{\rm short}\sim {\cal O}(1)$ in $\tau =10$ms using the mechanical frequency $\omega_m = 2\pi \times 1$ Hz and setting
the delocalization to $\Delta x=x_{\rm zpf}e^{\xi}\sim R \sim 1$cm (corresponding to a squeezing parameter $\xi=35$) while maintaining the same laser intensity. We also note that a small change in the linear entropy resulting from prograde and retrograde motion with respect to the spinning sphere is a measurable quantity in a laboratory. A rough estimation for an $M=10$ kg highly-dense sphere with a high breaking stress, such as in the case of Iridium, can provide $S_{J,+}^{\rm short} -S_{J,-}^{\rm short}\sim 10^{-7}$. Although there is a small difference, this can be measured with an appropriate witnessing scheme. In the future, we will be keen to study the feasibility of the scheme in the experimental context and report a viable witnessing scheme.

\begin{acknowledgments}
SB would like to acknowledge EPSRC grants
(EP/N031105/1, EP/S000267/1, and EP/X009467/1)
and grant ST/W006227/1.
 S.B and A.M.'s research are funded by the Gordon and Betty Moore Foundation through Grant GBMF12328, DOI 10.37807/GBMF12328. This material is based on work supported by the Alfred P. Sloan Foundation under Grant No. G-2023-21130. M.T. acknowledges funding from the Slovenian Research and Innovation Agency (ARIS) under Contracts No. N1-0392, No. P1-0416, and No. SN-ZRD/22-27/0510 (RSUL Toroš). 
\end{acknowledgments}

\bibliography{Ref.bib}

@article{Dyson:1920cwa,
    author = "Dyson, F. W. and Eddington, A. S. and Davidson, C.",
    title = "{A Determination of the Deflection of Light by the Sun's Gravitational Field, from Observations Made at the Total Eclipse of May 29, 1919}",
    doi = "10.1098/rsta.1920.0009",
    journal = "Phil. Trans. Roy. Soc. Lond. A",
    volume = "220",
    pages = "291--333",
    year = "1920"
}

@article{Rufo:2024ulr,
    author = "Rufo, Pablo Guillermo Carmona and Mazumdar, Anupam and Sab{\'\i}n, Carlos",
    title = "{Genuine tripartite entanglement in graviton-matter interactions}",
    eprint = "2411.03293",
    archivePrefix = "arXiv",
    primaryClass = "quant-ph",
    doi = "10.1103/PhysRevA.111.022444",
    journal = "Phys. Rev. A",
    volume = "111",
    number = "2",
    pages = "022444",
    year = "2025"
}

@article{Dutta:2026pty,
  title={Witnessing entanglement between photon and matter due to graviton exchange},
  author={Dutta, Arijit and Toro{\v{s}}, Marko and Bose, Sougato and Mazumdar, Anupam},
  journal={arXiv preprint arXiv:2604.24496},
  year={2026}
}

@article{Bose:2017nin,
    author = "Bose, Sougato and Mazumdar, Anupam and Morley, Gavin W. and Ulbricht, Hendrik and Toro\v{s}, Marko and Paternostro, Mauro and Geraci, Andrew and Barker, Peter and Kim, M. S. and Milburn, Gerard",
    title = "{Spin Entanglement Witness for Quantum Gravity}",
    eprint = "1707.06050",
    archivePrefix = "arXiv",
    primaryClass = "quant-ph",
    doi = "10.1103/PhysRevLett.119.240401",
    journal = "Phys. Rev. Lett.",
    volume = "119",
    number = "24",
    pages = "240401",
    year = "2017"
}

@article{Schut:2024lgp,
    author = "Schut, Martine and Andriolo, Patrick and Toro{\v{s}}, Marko and Bose, Sougato and Mazumdar, Anupam",
    title = "{Expression for the decoherence rate due to air-molecule scattering in spatial qubits}",
    eprint = "2410.20910",
    archivePrefix = "arXiv",
    primaryClass = "quant-ph",
    doi = "10.1103/PhysRevA.111.042211",
    journal = "Phys. Rev. A",
    volume = "111",
    number = "4",
    pages = "042211",
    year = "2025"
}

@article{Zhou:2024pdl,
    author = "Zhou, Tian and Bose, Sougato and Mazumdar, Anupam",
    title = "{Gyroscopic stability for nanoparticles in Stern-Gerlach Interferometry and spin contrast}",
    eprint = "2407.15813",
    archivePrefix = "arXiv",
    primaryClass = "quant-ph",
    doi = "10.1103/4xvz-gnk7",
    journal = "Phys. Rev. A",
    volume = "112",
    number = "1",
    pages = "013315",
    year = "2025"
}

@article{Krisnanda_2020,
   title={Observable quantum entanglement due to gravity},
   volume={6},
   ISSN={2056-6387},
   url={http://dx.doi.org/10.1038/s41534-020-0243-y},
   DOI={10.1038/s41534-020-0243-y},
   number={1},
   journal={npj Quantum Information},
   publisher={Springer Science and Business Media LLC},
   author={Krisnanda, Tanjung and Tham, Guo Yao and Paternostro, Mauro and Paterek, Tomasz},
   year={2020},
   month=Jan }

@article{Marshman:2021wyk,
    author = "Marshman, Ryan J. and Mazumdar, Anupam and Folman, Ron and Bose, Sougato",
    title = "{Constructing nano-object quantum superpositions with a Stern-Gerlach interferometer}",
    eprint = "2105.01094",
    archivePrefix = "arXiv",
    primaryClass = "quant-ph",
    doi = "10.1103/PhysRevResearch.4.023087",
    journal = "Phys. Rev. Res.",
    volume = "4",
    number = "2",
    pages = "023087",
    year = "2022"
}

@article{Tilly:2021qef,
    author = "Tilly, Jules and Marshman, Ryan J. and Mazumdar, Anupam and Bose, Sougato",
    title = "{Qudits for witnessing quantum-gravity-induced entanglement of masses under decoherence}",
    eprint = "2101.08086",
    archivePrefix = "arXiv",
    primaryClass = "quant-ph",
    doi = "10.1103/PhysRevA.104.052416",
    journal = "Phys. Rev. A",
    volume = "104",
    number = "5",
    pages = "052416",
    year = "2021"
}

@article{Saulson:1984,
  title = {Terrestrial gravitational noise on a gravitational wave antenna},
  author = {Saulson, Peter R.},
  journal = {Phys. Rev. D},
  volume = {30},
  issue = {4},
  pages = {732--736},
  numpages = {0},
  year = {1984},
  month = {Aug},
  publisher = {American Physical Society},
  doi = {10.1103/PhysRevD.30.732},
  url = {https://link.aps.org/doi/10.1103/PhysRevD.30.732}
}

@misc{ICTS, 
author = {},
title = {},
howpublished = "\url{https://www.youtube.com/watch?v=0Fv-0k13s_k}",
year = {2016}, 
note = "Accessed 1/11/22",
}

@article{Marshman:2019sne,
    author = "Marshman, Ryan J. and Mazumdar, Anupam and Bose, Sougato",
    title = "{Locality and entanglement in table-top testing of the quantum nature of linearized gravity}",
    eprint = "1907.01568",
    archivePrefix = "arXiv",
    primaryClass = "quant-ph",
    doi = "10.1103/PhysRevA.101.052110",
    journal = "Phys. Rev. A",
    volume = "101",
    number = "5",
    pages = "052110",
    year = "2020"
}

@article{Bose:2022uxe,
    author = "Bose, Sougato and Mazumdar, Anupam and Schut, Martine and Toro\v{s}, Marko",
    title = "{Mechanism for the quantum natured gravitons to entangle masses}",
    eprint = "2201.03583",
    archivePrefix = "arXiv",
    primaryClass = "gr-qc",
    doi = "10.1103/PhysRevD.105.106028",
    journal = "Phys. Rev. D",
    volume = "105",
    number = "10",
    pages = "106028",
    year = "2022"
}

@article{elahi2023probing,
  title={Probing massless and massive gravitons via entanglement in a warped extra dimension},
  author={Elahi, Shafaq Gulzar and Mazumdar, Anupam},
  doi="10.1103/PhysRevD.108.035018",
  journal={Physical Review D},
  volume={108},
  number={3},
  pages={035018},
  year={2023},
  publisher={APS}
}

@article{Vinckers:2023grv,
  title={Quantum entanglement of masses with nonlocal gravitational interaction},
  author={Beckering Vinckers, Ulrich K and De La Cruz-Dombriz, {\'A}lvaro and Mazumdar, Anupam},
  doi="10.1103/PhysRevD.107.124036",
  journal={Physical Review D},
  volume={107},
  number={12},
  pages={124036},
  year={2023},
  publisher={APS}
}

@article{Marletto:2017kzi,
    author = "Marletto, Chiara and Vedral, Vlatko",
    title = "{Gravitationally-induced entanglement between two massive particles is sufficient evidence of quantum effects in gravity}",
    eprint = "1707.06036",
    archivePrefix = "arXiv",
    primaryClass = "quant-ph",
    doi = "10.1103/PhysRevLett.119.240402",
    journal = "Phys. Rev. Lett.",
    volume = "119",
    number = "24",
    pages = "240402",
    year = "2017"
}

@article{Toros:2020dbf,
    author = "Toro\v{s}, Marko and Van De Kamp, Thomas W. and Marshman, Ryan J. and Kim, M. S. and Mazumdar, Anupam and Bose, Sougato",
    title = "{Relative acceleration noise mitigation for nanocrystal matter-wave interferometry: Applications to entangling masses via quantum gravity}",
    eprint = "2007.15029",
    archivePrefix = "arXiv",
    primaryClass = "gr-qc",
    doi = "10.1103/PhysRevResearch.3.023178",
    journal = "Phys. Rev. Res.",
    volume = "3",
    number = "2",
    pages = "023178",
    year = "2021"
}

@article{vandeKamp:2020rqh,
    author = "van de Kamp, Thomas W. and Marshman, Ryan J. and Bose, Sougato and Mazumdar, Anupam",
    title = "{Quantum Gravity Witness via Entanglement of Masses: Casimir Screening}",
    eprint = "2006.06931",
    archivePrefix = "arXiv",
    primaryClass = "quant-ph",
    doi = "10.1103/PhysRevA.102.062807",
    journal = "Phys. Rev. A",
    volume = "102",
    number = "6",
    pages = "062807",
    year = "2020"
}

@article{Toros:2020krn,
  title={Loss of coherence and coherence protection from a graviton bath},
  author={Toro{\v{s}}, Marko and Mazumdar, Anupam and Bose, Sougato},
  doi="10.1103/PhysRevD.109.084050",
  journal={Physical Review D},
  volume={109},
  number={8},
  pages={084050},
  year={2024},
  publisher={APS}
}

@article{Chevalier:2020uvv,
    author = "Chevalier, Hadrien and Paige, A. J. and Kim, M. S.",
    title = "{Witnessing the nonclassical nature of gravity in the presence of unknown interactions}",
    eprint = "2005.13922",
    archivePrefix = "arXiv",
    primaryClass = "quant-ph",
    doi = "10.1103/PhysRevA.102.022428",
    journal = "Phys. Rev. A",
    volume = "102",
    number = "2",
    pages = "022428",
    year = "2020"
}

@article{Carney_2019,
doi = {10.1088/1361-6382/aaf9ca},
url = {https://dx.doi.org/10.1088/1361-6382/aaf9ca},
year = {2019},
publisher = {IOP Publishing},
volume = {36},

pages = {034001},
author = {Daniel Carney and Philip C E Stamp and Jacob M Taylor},
title = {Tabletop experiments for quantum gravity: a user's manual},
journal = {Class. Quant. Grav.},
}

@article{Belenchia:2018szb,
    author = "Belenchia, Alessio and others",
    title = "{Quantum Superposition of Massive Objects and the Quantization of Gravity}",
    
    doi = "10.1103/PhysRevD.98.126009",
    journal = "Phys. Rev. D",
    volume = "98",
   
    pages = "126009",
    year = "2018"
}

@article{Carney:2021vvt,
    author = "Carney, Daniel",
    title = "{Newton, entanglement, and the graviton}",
    eprint = "2108.06320",
    archivePrefix = "arXiv",
    primaryClass = "quant-ph",
    doi = "10.1103/PhysRevD.105.024029",
    journal = "Phys. Rev. D",
    volume = "105",
    number = "2",
    pages = "024029",
    year = "2022"
}

@article{Danielson:2021egj,
    author = "Danielson, Daine L. and Satishchandran, Gautam and Wald, Robert M.",
    title = "{Gravitationally mediated entanglement: Newtonian field versus gravitons}",
    
    doi = "10.1103/PhysRevD.105.086001",
    journal = "Phys. Rev. D",
    volume = "105",
    
    pages = "086001",
    year = "2022"
}

@inproceedings{Bose:2025qns,
    author = "Bose, Sougato and others",
    title = "{A Spin-Based Pathway to Testing the Quantum Nature of Gravity}",
    eprint = "2509.01586",
    archivePrefix = "arXiv",
    primaryClass = "quant-ph",
    month = "9",
    year = "2025"
}

@inproceedings{Biswas:2013kla,
    author = "Biswas, Tirthabir and Koivisto, Tomi and Mazumdar, Anupam",
    title = "{Nonlocal theories of gravity: the flat space propagator}",
    booktitle = "{Barcelona Postgrad Encounters on Fundamental Physics}",
    eprint = "1302.0532",
    archivePrefix = "arXiv",
    primaryClass = "gr-qc",
    pages = "13--24",
    year = "2013"
}

@article{Biswas:2011ar,
    author = "Biswas, Tirthabir and Gerwick, Erik and Koivisto, Tomi and Mazumdar, Anupam",
    title = "{Towards singularity and ghost free theories of gravity}",
    eprint = "1110.5249",
    archivePrefix = "arXiv",
    primaryClass = "gr-qc",
    doi = "10.1103/PhysRevLett.108.031101",
    journal = "Phys. Rev. Lett.",
    volume = "108",
    pages = "031101",
    year = "2012"
}

@article{christodoulou2023locally,
  title={Locally mediated entanglement in linearized quantum gravity},
  author={Christodoulou, Marios and Di Biagio, Andrea and Aspelmeyer, Markus and Brukner, {\v{C}}aslav and Rovelli, Carlo and Howl, Richard},
  doi = "10.1103/PhysRevLett.130.100202",
  journal={Physical Review Letters},
  volume={130},
  number={10},
  pages={100202},
  year={2023},
  publisher={APS}
}

@article{Elahi:2024dbb,
    author = "Elahi, Shafaq Gulzar and Schut, Martine and Dana, Andrew and Grinin, Alexey and Bose, Sougato and Mazumdar, Anupam and Geraci, Andrew",
    title = "{Diamagnetic micro-chip traps for levitated nanoparticle entanglement experiments}",
    journal={arXiv preprint arXiv:2411.02325},
    eprint = "2411.02325",
    archivePrefix = "arXiv",
    primaryClass = "quant-ph",
    month = "11",
    year = "2024"
}

@article{Perdriat:2024xiy,
    author = "Perdriat, Maxime and Rusconi, Cosimo C. and Delord, Tom and Huillery, Paul and Pellet-Mary, Cl\'ement and Durand, Alrik and Stickler, Benjamin A. and H\'etet, Gabriel",
    title = "{Rotational Locking of Charged Microparticles in Quadrupole Ion Traps}",
    doi = "10.1103/PhysRevLett.133.253602",
    journal = "Phys. Rev. Lett.",
    volume = "133",
    number = "25",
    pages = "253602",
    year = "2024"
}

@article{Perdriat:2024zje,
    author = "Perdriat, Maxime and Durand, Alrik and Voisin, Julien and H\'etet, Gabriel",
    title = "{Spin-dependent Force from an NV center Ensemble on a Microlever}",
    journal={arXiv preprint arXiv:2410.18762},
    eprint = "2410.18762",
    archivePrefix = "arXiv",
    primaryClass = "quant-ph",
    month = "10",
    year = "2024"
}

@article{Schlosshauer:2019ewh,
    author = "Schlosshauer, Maximilian",
    title = "{Quantum Decoherence}",
    eprint = "1911.06282",
    archivePrefix = "arXiv",
    primaryClass = "quant-ph",
    doi = "10.1016/j.physrep.2019.10.001",
    journal = "Phys. Rept.",
    volume = "831",
    pages = "1--57",
    year = "2019"
}

@article{Biswas:2022qto,
    author = "Biswas, Dripto and Bose, Sougato and Mazumdar, Anupam and Toro\v{s}, Marko",
    title = "{Gravitational optomechanics: Photon-matter entanglement via graviton exchange}",
    eprint = "2209.09273",
    archivePrefix = "arXiv",
    primaryClass = "gr-qc",
    doi = "10.1103/PhysRevD.108.064023",
    journal = "Phys. Rev. D",
    volume = "108",
    number = "6",
    pages = "064023",
    year = "2023"
}

@article{Kafri_2015,
doi = {10.1088/1367-2630/17/1/015006},
url = {https://dx.doi.org/10.1088/1367-2630/17/1/015006},
year = {2015},
month = {jan},
publisher = {IOP Publishing},
volume = {17},
number = {1},
pages = {015006},
author = {D Kafri and G J Milburn and J M Taylor},
title = {Bounds on quantum communication via Newtonian gravity},
journal = {New Journal of Physics}
}

@article{Gupta,
doi = {10.1088/0370-1298/65/8/304},
url = {https://dx.doi.org/10.1088/0370-1298/65/8/304},
year = {1952},
month = {aug},
publisher = {},
volume = {65},
number = {8},
pages = {608},
author = {Suraj N Gupta},
title = {Quantization of Einstein's Gravitational Field: General Treatment},
journal = {Proceedings of the Physical Society. Section A}
}

@article{Donoghue:1994dn,
    author = "Donoghue, John F.",
    title = "{General relativity as an effective field theory: The leading quantum corrections}",
    eprint = "gr-qc/9405057",
    archivePrefix = "arXiv",
    reportNumber = "UMHEP-408",
    doi = "10.1103/PhysRevD.50.3874",
    journal = "Phys. Rev. D",
    volume = "50",
    pages = "3874--3888",
    year = "1994"
}

@book{Scadron:2007qd,
    title={Advanced Quantum Theory},
    author={Scadron, Michael D.},
    year={2007},
    edition={3rd},
    publisher={World Scientific},
    address={Singapore},
    isbn={9789812700506},
}

@article{Pedernales20_GM,
   author = {Pedernales, Julen S. and Morley, Gavin W. and Plenio, Martin B.},
   title = {Motional Dynamical Decoupling for Interferometry with Macroscopic Particles},
   journal = {Phys. Rev. Lett.},
   volume = {125},
 
   pages = {023602},
   DOI = {10.1103/PhysRevLett.125.023602},
   url = {https://link.aps.org/doi/10.1103/PhysRevLett.125.023602},
   year = {2020},
   type = {Journal Article}
}

@article{ORI11_GM,
   author = {Romero-Isart, Oriol},
   title = {Quantum superposition of massive objects and collapse models},
   journal = {Phys. Rev. A},
   volume = {84},  
   pages = {052121},
   DOI = {10.1103/PhysRevA.84.052121},
   url = {https://link.aps.org/doi/10.1103/PhysRevA.84.052121},
   year = {2011},
   type = {Journal Article}
}

@article{Hughes_1998,
   title={Seismic gravity-gradient noise in interferometric gravitational-wave detectors},
   volume={58},
   ISSN={1089-4918},
   url={http://dx.doi.org/10.1103/PhysRevD.58.122002},
   DOI={10.1103/physrevd.58.122002},
   number={12},
   journal={Physical Review D},
   publisher={American Physical Society (APS)},
   author={Hughes, Scott A. and Thorne, Kip S.},
   year={1998},
   month=nov }

@article{Thorne_1999,
   title={Human gravity-gradient noise in interferometric gravitational-wave detectors},
   volume={60},
   ISSN={1089-4918},
   url={http://dx.doi.org/10.1103/PhysRevD.60.082001},
   DOI={10.1103/physrevd.60.082001},
   number={8},
   journal={Physical Review D},
   publisher={American Physical Society (APS)},
   author={Thorne, Kip S. and Winstein, Carolee J.},
   year={1999},
   month=sep }

@article{Stickler:2021dho,
    author = "Stickler, Benjamin A. and Hornberger, Klaus and Kim, M. S.",
    title = "{Quantum rotations of nanoparticles}",
    eprint = "2102.00992",
    archivePrefix = "arXiv",
    primaryClass = "quant-ph",
    doi = "10.1038/s42254-021-00335-0",
    journal = "Nature Rev. Phys.",
    volume = "3",
    pages = "589--597",
    year = "2021"
}

@article{Whittle_2021,
	doi = {10.1126/science.abh2634},
  
	url = {https://doi.org/10.1126%2Fscience.abh2634},
  
	year = 2021,
	month = {jun},
  
	publisher = {American Association for the Advancement of Science ({AAAS})},
  
	volume = {372},
  
	number = {6548},
  
	pages = {1333--1336},
  
	author = {Chris Whittle},
  
	title = {Approaching the motional ground state of a 10-kg object},
  
	journal = {et.al., Science}
}

@ARTICLE{Bodenner:2003,
       author = {{Bodenner}, Jeremiah and {Will}, Clifford M.},
        title = "{Deflection of light to second order: A tool for illustrating principles of general relativity}",
      journal = {American Journal of Physics},
         year = 2003,
        month = aug,
       volume = {71},
       number = {8},
        pages = {770-773},
          doi = {10.1119/1.1570416},
       adsurl = {https://ui.adsabs.harvard.edu/abs/2003AmJPh..71..770B}
}

@article{Toros:2020,
   title={Revealing and concealing entanglement with noninertial motion},
   volume={101},
   ISSN={2469-9934},
   url={http://dx.doi.org/10.1103/PhysRevA.101.043837},
   DOI={10.1103/physreva.101.043837},
   number={4},
   journal={Physical Review A},
   publisher={American Physical Society (APS)},
   author={Toroš, Marko and Restuccia, Sara and Gibson, Graham M. and Cromb, Marion and Ulbricht, Hendrik and Padgett, Miles and Faccio, Daniele},
   year={2020},
   month=Apr }

@article{PhysRevLett.123.110401,
  title = {Photon Bunching in a Rotating Reference Frame},
  author = {Restuccia, Sara and Toro\ifmmode \check{s}\else \v{s}\fi{}, Marko and Gibson, Graham M. and Ulbricht, Hendrik and Faccio, Daniele and Padgett, Miles J.},
  journal = {Phys. Rev. Lett.},
  volume = {123},
  issue = {11},
  pages = {110401},
  numpages = {6},
  year = {2019},
  month = {Sep},
  publisher = {American Physical Society},
  doi = {10.1103/PhysRevLett.123.110401},
  url = {https://link.aps.org/doi/10.1103/PhysRevLett.123.110401}
}

@article{Fink_2017,
   title={Experimental test of photonic entanglement in accelerated reference frames},
   volume={8},
   ISSN={2041-1723},
   url={http://dx.doi.org/10.1038/ncomms15304},
   DOI={10.1038/ncomms15304},
   number={1},
   journal={Nature Communications},
   publisher={Springer Science and Business Media LLC},
   author={Fink, Matthias and Rodriguez-Aramendia, Ana and Handsteiner, Johannes and Ziarkash, Abdul and Steinlechner, Fabian and Scheidl, Thomas and Fuentes, Ivette and Pienaar, Jacques and Ralph, Timothy C. and Ursin, Rupert},
   year={2017},
   month=May }

@article{Toros:2022,
   title={Generation of Entanglement from Mechanical Rotation},
   volume={129},
   ISSN={1079-7114},
   url={http://dx.doi.org/10.1103/PhysRevLett.129.260401},
   DOI={10.1103/physrevlett.129.260401},
   number={26},
   journal={Physical Review Letters},
   publisher={American Physical Society (APS)},
   author={Toroš, Marko and Cromb, Marion and Paternostro, Mauro and Faccio, Daniele},
   year={2022},
   month=Dec }

@ARTICLE{Yoon:2021,
       author = {{Yoon}, Jin Woo and {Kim}, Yeong Gyu and {Choi}, Il Woo and {Sung}, Jae Hee and {Lee}, Hwang Woon and {Lee}, Seong Ku and {Nam}, Chang Hee},
        title = "{Realization of laser intensity over 1023 W/cm2}",
      journal = {Optica},
         year = 2021,
        month = may,
       volume = {8},
       number = {5},
        pages = {630},
          doi = {10.1364/OPTICA.420520},
       adsurl = {https://ui.adsabs.harvard.edu/abs/2021Optic...8..630Y}
}

@article{vovrosh2017parametric,
  title={Parametric feedback cooling of levitated optomechanics in a parabolic mirror trap},
  author={Vovrosh, Jamie and Rashid, Muddassar and Hempston, David and Bateman, James and Paternostro, Mauro and Ulbricht, Hendrik},
  journal={JOSA B},
  volume={34},
  number={7},
  pages={1421--1428},
  year={2017},
  publisher={Optica Publishing Group}
}

@article{Ahn:2018,
  title = {Optically Levitated Nanodumbbell Torsion Balance and GHz Nanomechanical Rotor},
  author = {Ahn, Jonghoon and Xu, Zhujing and Bang, Jaehoon and Deng, Yu-Hao and Hoang, Thai M. and Han, Qinkai and Ma, Ren-Min and Li, Tongcang},
  journal = {Phys. Rev. Lett.},
  volume = {121},
  issue = {3},
  pages = {033603},
  numpages = {5},
  year = {2018},
  month = {Jul},
  publisher = {American Physical Society},
  doi = {10.1103/PhysRevLett.121.033603},
  url = {https://link.aps.org/doi/10.1103/PhysRevLett.121.033603}
}

@article{
Schuck:2018,
author = {Marcel Schuck  and Daniel Steinert  and Thomas Nussbaumer  and Johann W. Kolar },
title = {Ultrafast rotation of magnetically levitated macroscopic steel spheres},
journal = {Science Advances},
volume = {4},
number = {1},
pages = {e1701519},
year = {2018},
doi = {10.1126/sciadv.1701519},
URL = {https://www.science.org/doi/abs/10.1126/sciadv.1701519},
eprint = {https://www.science.org/doi/pdf/10.1126/sciadv.1701519}}

@article{Hornberger:2008xkz,
    author = "Hornberger, Klaus",
    title = "{Introduction to decoherence theory}",
    eprint = "quant-ph/0612118",
    archivePrefix = "arXiv",
    doi = "10.1007/978-3-540-88169-8_5",
    journal = "Lect. Notes Phys.",
    volume = "768",
    pages = "221--276",
    year = "2009"
}

@article{caves1981quantum,
  title     = {Quantum-mechanical noise in an interferometer},
  author    = {Caves, Carlton M.},
  journal   = {Physical Review D},
  volume    = {23},
  number    = {8},
  pages     = {1693--1708},
  year      = {1981},
  publisher = {APS},
  doi       = {10.1103/PhysRevD.23.1693}
}

@misc{Preskill-noise,
  title = {Notes on noise},
  author={J. Preskill},
  howpublished = {http://theory.caltech.edu/~preskill/papers/decoherence\_notesv2.pdf},
}

@article{christodoulou2019possibility,
  title={On the possibility of laboratory evidence for quantum superposition of geometries},
  author={Christodoulou, Marios and Rovelli, Carlo},
  doi="10.1016/j.physletb.2019.03.015",
  journal={Physics Letters B},
  volume={792},
  pages={64--68},
  year={2019},
  publisher={Elsevier}
}

@article{Kopeikin:2001dz,
    author = "Kopeikin, Sergei and Mashhoon, Bahram",
    title = "{Gravitomagnetic effects in the propagation of electromagnetic waves in variable gravitational fields of arbitrary moving and spinning bodies}",
    eprint = "gr-qc/0110101",
    archivePrefix = "arXiv",
    doi = "10.1103/PhysRevD.65.064025",
    journal = "Phys. Rev. D",
    volume = "65",
    pages = "064025",
    year = "2002"
}

@article{Guadagnini:2002xx,
    author = "Guadagnini, Enore",
    title = "{Gravitational deflection of light and helicity asymmetry}",
    eprint = "gr-qc/0207036",
    archivePrefix = "arXiv",
    reportNumber = "IFUP-TH-2002-26",
    doi = "10.1016/S0370-2693(02)02811-3",
    journal = "Phys. Lett. B",
    volume = "548",
    pages = "19--23",
    year = "2002"
}
\end{document}